\input harvmac.tex

\newread\epsffilein    
\newif\ifepsffileok    
\newif\ifepsfbbfound   
\newif\ifepsfverbose   
\newdimen\epsfxsize    
\newdimen\epsfysize    
\newdimen\epsftsize    
\newdimen\epsfrsize    
\newdimen\epsftmp      
\newdimen\pspoints     
\pspoints=1bp          
\epsfxsize=0pt         
\epsfysize=0pt         
\def\epsfbox#1{\global\def\epsfllx{72}\global\def\epsflly{72}%
   \global\def\epsfurx{540}\global\def\epsfury{720}%
   \def\lbracket{[}\def\testit{#1}\ifx\testit\lbracket
   \let\next=\epsfgetlitbb\else\let\next=\epsfnormal\fi\next{#1}}%
\def\epsfgetlitbb#1#2 #3 #4 #5]#6{\epsfgrab #2 #3 #4 #5 .\\%
   \epsfsetgraph{#6}}%
\def\epsfnormal#1{\epsfgetbb{#1}\epsfsetgraph{#1}}%
\def\epsfgetbb#1{%
%
%
\openin\epsffilein=#1
\ifeof\epsffilein\errmessage{I couldn't open #1, will ignore it}\else
%
%
   {\epsffileoktrue \chardef\other=12
    \def\do##1{\catcode`##1=\other}\dospecials \catcode`\ =10
    \loop
       \read\epsffilein to \epsffileline
       \ifeof\epsffilein\epsffileokfalse\else
%
%
          \expandafter\epsfaux\epsffileline:. \\%
       \fi
   \ifepsffileok\repeat
   \ifepsfbbfound\else
    \ifepsfverbose\message{No bounding box comment in #1; using defaults}\fi\fi
   }\closein\epsffilein\fi}%
%
%
\def\epsfclipstring{}
\def\epsfsetgraph#1{%
   \epsfrsize=\epsfury\pspoints
   \advance\epsfrsize by-\epsflly\pspoints
   \epsftsize=\epsfurx\pspoints
   \advance\epsftsize by-\epsfllx\pspoints
%
%
   \epsfxsize\epsfsize\epsftsize\epsfrsize
   \ifnum\epsfxsize=0 \ifnum\epsfysize=0
      \epsfxsize=\epsftsize \epsfysize=\epsfrsize
      \epsfrsize=0pt
%
%
     \else\epsftmp=\epsftsize \divide\epsftmp\epsfrsize
       \epsfxsize=\epsfysize \multiply\epsfxsize\epsftmp
       \multiply\epsftmp\epsfrsize \advance\epsftsize-\epsftmp
       \epsftmp=\epsfysize
       \loop \advance\epsftsize\epsftsize \divide\epsftmp 2
       \ifnum\epsftmp>0
          \ifnum\epsftsize<\epsfrsize\else
             \advance\epsftsize-\epsfrsize \advance\epsfxsize\epsftmp \fi
       \repeat
       \epsfrsize=0pt
     \fi
   \else \ifnum\epsfysize=0
     \epsftmp=\epsfrsize \divide\epsftmp\epsftsize
     \epsfysize=\epsfxsize \multiply\epsfysize\epsftmp   
     \multiply\epsftmp\epsftsize \advance\epsfrsize-\epsftmp
     \epsftmp=\epsfxsize
     \loop \advance\epsfrsize\epsfrsize \divide\epsftmp 2
     \ifnum\epsftmp>0
        \ifnum\epsfrsize<\epsftsize\else
           \advance\epsfrsize-\epsftsize \advance\epsfysize\epsftmp \fi
     \repeat
     \epsfrsize=0pt
    \else
     \epsfrsize=\epsfysize
    \fi
   \fi
%
%
   \ifepsfverbose\message{#1: width=\the\epsfxsize, height=\the\epsfysize}\fi
   \epsftmp=10\epsfxsize \divide\epsftmp\pspoints
   \vbox to\epsfysize{\vfil\hbox to\epsfxsize{%
      \ifnum\epsfrsize=0\relax
        \includegraphics{#1}%
      \else
        \epsfrsize=10\epsfysize \divide\epsfrsize\pspoints
        \includegraphics{#1}%
      \fi
      \hfil}}%
\global\epsfxsize=0pt\global\epsfysize=0pt}%
%
%
{\catcode`\%=12 \global\let\epsfpercent=
%
%
\long\def\epsfaux#1#2:#3\\{\ifx#1\epsfpercent
   \def\testit{#2}\ifx\testit\epsfbblit
      \epsfgrab #3 . . . \\%
      \epsffileokfalse
      \global\epsfbbfoundtrue
   \fi\else\ifx#1\par\else\epsffileokfalse\fi\fi}%
%
%
\def\epsfempty{}%
\def\epsfgrab #1 #2 #3 #4 #5\\{%
\global\def\epsfllx{#1}\ifx\epsfllx\epsfempty
      \epsfgrab #2 #3 #4 #5 .\\\else
   \global\def\epsflly{#2}%
   \global\def\epsfurx{#3}\global\def\epsfury{#4}\fi}%
%
%
\def\epsfsize#1#2{\epsfxsize}
%
%

%
\ifx\epsfbox\UnDeFiNeD\message{(NO epsf.tex, FIGURES WILL BE
IGNORED)}
\def\Fig.in#1{\vskip2in}
\else\message{(FIGURES WILL BE INCLUDED)}\def\Fig.in#1{#1}\fi
\def\iFig.#1#2#3{\xdef#1{Fig..~\the\Fig.no}
\goodbreak\topinsert\Fig.in{\centerline{#3}}%
\smallskip\centerline{\vbox{\baselineskip12pt
\advance\hsize by -1truein\noindent{\bf Fig..~\the\Fig.no:} #2}}
\bigskip\endinsert\global\advance\Fig.no by1}

\def\cst {{\rm const.}}

\def \ov {\over}

\def \lr { \lref}

\def\dd {\partial }

\def\l{\lambda}

\def \k {\kappa}

\def\n{\noindent}
\gdef \jnl#1, #2, #3, 1#4#5#6{ { #1~}{ #2} (1#4#5#6) #3}

\def \prl { Phys. Rev. Lett. }
\def \pr  { Phys. Rev. }
\def \cqg { Class. Quant. Grav. }

\lr \blgu {S.K. Blau and A.H. Guth: {\it Inflationary Cosmology}
in `{\it 300 years of Gravitation}' pag. 524, ed. by S.W. Hawking 
and W. Israel (Cambridge University Press, 1987).}
\lr \cghs {C. Callan, S. Giddings, J. Harvey and A. Strominger,
\pr D45 (1992) R1005. }
\lr \birdav {N.D. Birrell and P.C.W. Davies, {\it Quantum fields in
curved space} (Cambridge University Press, Cambridge, England, 1982). }
\lr \schutz {B.F. Schutz, `` A first course in general relativity",
(Cambridge University Press, Cambridge, England, 1985).}
\lr \poisson{E. Poisson, private communication.}
\lr \rst {J.G. Russo, L. Susskind and L. Thorlacius,
\pr D46 (1992) 3444; \pr D47 (1993) 533.  }
\lr \hawk {S. Hawking, Commun. Math. Phys. 43 (1975) 199.  }
\lr\wald {R. M. Wald, {\it General Relativity} (University of
Chicago Press, Chicago, 1984).}
\lr\wein{S. Weinberg, {\it Gravitation and Cosmology},
John Wiley, Inc., New York (1972).}
\lr\trivedi{S.P. Trivedi, \pr D47 (1993) 4233.}
\lr \cghs {C. Callan, S. Giddings, J. Harvey and A. Strominger, \pr\ D45 (1992)
R1005.}
\lr \fabru {A. Fabbri and J.G. Russo, \pr D53 (1996) 6995.}
\lr \balfab {R. Balbinot and A. Fabbri, gr-qc/9602047 (to appear in \cqg).}
\lr \fdu {W.G. Unruh, \pr\ D14 (1976) 870.}
\lr \penrose {R. Penrose in {\it Battelle Rencontres}, ed. by C.M. De Witt
and J.A. Wheeler (Benjamin, New York, 1968).}
\lr \mainf {E. Poisson and W. Israel,
\pr\ D41 (1990) 1796; A. Ori, \prl\ 67 (1991) 789.}
\lr \balb {R. Balbinot and P.R. Brady, \cqg\ 11 (1994) 1763.}
\lr \price {R.H. Price, \pr\ D5 (1972) 2419.}
\lr \hawkelli {S.W. Hawking and G.F.R. Ellis, {\it The large scale
structure of space-time} 
(Cambridge University Press, Cambridge, England, 1973). }
\lr \wald {R. M. Wald, {\it General Relativity} (University of
Chicago Press, Chicago, 1984).} 
\lr \chma {J.S.F. Chan and R.B. Mann, \pr D50 (1994) 7376.}
\lr \bradpoi {P.R. Brady and E. Poisson, \cqg\ 9 (1992) 121.}
\lr \branusi {P.R. Brady, D. Nunez and S. Sinha, \pr\ D47 (1993)
 4239.}
\lr \poimar {D. Markovic and E. Poisson, \prl\ 74 (1995) 1280.}
\lr \fabal {R. Balbinot and A. Fabbri, {\it Two-dimensional black holes
in accelerated frames: quantum aspects}, to appear}

\baselineskip8pt
\Title{\vbox
{\baselineskip 6pt 
\hbox{SISSA-ISAS/102/96/EP} 
{\hbox{
   }}} }
{\vbox{\centerline { Classical stability of black hole Cauchy horizons} 
\vskip .2in \centerline {in two-dimensional asymptotically
flat space-times}
}}
\bigskip\bigskip\bigskip
\vskip -20 true pt
\centerline { A. Fabbri }
\smallskip \bigskip
\centerline {\it SISSA-ISAS and INFN sezione di Trieste,  }
\smallskip
\centerline {\it Via Beirut 2-4, 34014 Trieste, Italy}
\smallskip
\centerline {\it   fabbri@gandalf.sissa.it}
\bigskip\bigskip\bigskip
\bigskip\bigskip\bigskip
\centerline {\bf Abstract}
\bigskip
In this paper we analyse the stability of black hole
Cauchy horizons arising in a class of 2d dilaton gravity models.
\par
It is shown that due to the characteristic asymptotic Rindler form of the 
metric of these models,
time dependent gravitational perturbations generated in the external
region do not necessarily blow up when propagated
along the Cauchy horizon. There exists, in fact, 
a region of nonzero measure in the space of the parameters 
characterizing the solutions such that both instability and
mass inflation are avoided.
\par
This is a new result concerning asymptotically flat space-times,
not shared by the well-known solutions of General Relativity.
\par
Despite this fact, however, quantum backreaction seems to produce
a scalar curvature singularity there.

\medskip
\baselineskip8pt
\noindent

\Date {July 1996}

\noblackbox
\baselineskip 14pt plus 2pt minus 2pt

\vfill\eject

\newsec{ Introduction}
In the recent years there has been a long debate on the question of
whether Einstein field equations alone are able to completely
determine the evolution of the gravitational field, once initial
data are given on a Cauchy hypersurface $\Sigma$.
\par
There exist, in fact, spacetimes possessing null surfaces called 
Cauchy horizons - boundaries beyond which the future evolution of 
physical fields is no longer uniquely determinable from the data
prescribed on $\Sigma$.
Extra boundary conditions are then needed, sometimes, at spacetime
singularities and of course these are completely arbitrary
in the context of
the classical theory.
It seems, therefore, that predictability is lost at the Cauchy
horizons in Einstein gravity (see for example Ref. \hawkelli).
\par
We will divide the geometries of our interest endowed with Cauchy horizons in
two classes depending on their asymptotics; they are either
\par \n
i) asymptotically flat
\par \n
or
\par \n
ii) non asymptotically flat.
\par 
The exact solutions of General Relativity included in i) are
all those of Kerr-Newmann type, with nonvanishing electric charge $Q$ 
and/or angular momentum.
A good representative of this class, that keeps
all the essential features we wish to discuss, is the
Reissner-Nordstr\"om solution, representing the spacetime 
of a spherically symmetric black hole of mass $m_0$ carrying a conserved 
abelian charge $Q$.
\par
The simplest way to have a non flat asymptotic region is to add
in the Einstein action a cosmological constant $\Lambda$.
As $\Lambda >0$, this results in the well known De Sitter solution,
that is commonly used in cosmology to describe the inflationary
era of our Universe (see for example Ref. \blgu).
We will take as a representative
of class ii) the Reissner-Nordstr\"om-De Sitter geometry,
which incorporates the features of a charged black hole
immersed in an expanding universe.
\par
The question of the classical
precictability of the field equations can also be reformulated in this way:
are Cauchy horizons traversible in physically realistic
spacetimes?
\par
It was noted long ago by Penrose \penrose\ that
the Cauchy horizon of the Reissner-Nordstr\"om geometry
(for $m_0>|Q|$)
is an infinite blueshift surface. This means that a small
perturbation in the external region is seen infinitely
(exponentially) blueshifted by free falling observers
who cross this surface.
\par \n 
 For macroscopic black holes originated from the 
gravitational collapse of massive stars, these perturbations are caused 
by the radiative tail determined by Price \price.
Therefore, this tail is likely to modify the geometry
of the spacetime close to the Cauchy horizon.
\par \n
Poisson, Israel and Ori (see \mainf) have constructed a simple model
that mimics the realistic scenario of the gravitational collapse
and shown that the local mass  function and the
scalar curvature $R$ diverge at the Cauchy horizon: this is the mass inflation
phenomenon.
In this way they save the predictive power of the theory:
a singularity forms at the Cauchy horizon and any extension
of the geometry beyond it is meaningless!
\par
The peculiarity of the Reissner-Nordstr\"om-De Sitter geometry
is the presence, besides the black hole horizons, of a cosmological 
horizon, which divides the black hole region from the cosmological
one.
Perturbations, regular at the cosmological horizon, have 
been constructed for this spacetime and simple calculations
show that there's a region of nonzero measure in parameter
space ($m_0$, $Q$, $\Lambda$) for which the Cauchy horizon is stable 
and the mass inflation doesn't occur (\bradpoi\ and \branusi).  
\par
In this paper we check the stability issue for Cauchy horizons
in the context of a one-parameter class of simple 2d dilaton-gravity
models introduced in Ref. \fabru.
They are described by the action $S=S_n + S_{EM}$, where 
\eqn\accla{
S_n={1\ov{2\pi}}\int d^2 x \sqrt{-g}[ e^{-{2\ov n}\phi}
(R + {4\ov n}(\nabla\phi)^2) + 4\l^2 e^{-2\phi}],
}
and the explictiit form of $S_{EM}$ will be shown in section 3. 
$R$ is the scalar curvature associated to the two-dimensional
metric tensor $g_{ab}$, $\phi$ is the dilaton field.
In the case $n=1$, eq. \accla\ is the usual CGHS action 
\cghs.
\par
The motivation for studying such models is essentially
that they are exactly solvable at the semiclassical level and
contain vacuum as well as asymptotically flat solutions.
\par 
In Ref. \balfab\ it has been considered the global causal structure
of the corresponding static black hole solutions.
It is shown that for $n<1$ they are asymptotically flat, and
therefore of the type i). However, they are different from
the black holes of General Relativity for one main reason: the spacetime is 
asymptotically flat in Rindler coordinates and not, as usual, in
Minkowski ones.
 \par 
When $n>1$ the region where
the spacetime becomes flat is instead an horizon, an 
acceleration horizon, where the line element takes again
the Rindler form.
The metric can be analitically extended across this horizon and
a true asymptotic region doesn't seem to exist.
\par
The analysis of the stability of the Cauchy horizons
in these spacetimes leads to the following results:
\par \n
- for $n>1$ the behaviour  of the Cauchy horizon doesn't
differ significantly from the Reissner-Nordstr\"om-De Sitter  
case;
\par \n
- in the case $n<1$, surprisingly, there remains a region of nonzero 
measure in parameter space ($m_0$, $Q$, $\l$, $n$) for which the Cauchy
horizon is stable and the mass inflation doesn't take place.
\par \n
There is a simple physical reason behind this result: 
at the Cauchy horizon
the (infinite)
 blueshift factor 
 is no more exponential but
  power-law in the external inertial time coordinate.
We will clarify on this point in section 4.
\par 
However, quantum backreaction seems to
alter these conclusions. Such an analysis has already been carried out
in the RST model in \balb\ and similar
results are obtained here, suggesting that a scalar curvature singularity
always forms at the Cauchy horizon, thus forbidding any further extension
of the geometry above it.   

\newsec{The Cauchy horizons in General
Relativity}
\subsec{The Reissner-Nordstr\"om spacetime}
Let us consider the Reissner-Nordstr\"om
spacetime. The
metric reads 
\eqn\ago{
ds^2 =-f(r)dt^2 + {{dr^2}\ov{f(r)}}+ r^2 d\Omega^2,
}
where  
\eqn\upa{
f=1-{{2m_0}\ov{r}}+{{Q^2}\ov{r^2}}
}
and $d\Omega^2$ is the metric on the unit two-sphere.
\par
The zeros of $f$, namely $r_{\pm}=m_0\pm \sqrt{m_0^2 - Q^2}$ for
 $m_0>|Q|$, 
represent the
Killing horizons. $r_+$ is the event horizon and $r_-$ the inner, and
 Cauchy, 
horizon.
\par
We define, for future use, the surface gravity at both horizons
\eqn\hori{
k_{\pm}\equiv {1\ov 2} |\dd_r f|_{r_{\pm}}={{\sqrt{m^2 - Q^2}}\ov{r_{\pm}^2}}.
}
$r=0$ is the location of the singularity 
and $r=\infty$ defines the flat
asymptotic region, where the spacetime becomes minkowskian.
The Penrose diagram of this spacetime is well known and is represented in 
Fig. 1. The null directions are represented by the coordinates $u$ and
$v$, defined by $dv=dt+{{d\sigma}\ov{f}}$, $du=dt-{{d\sigma}\ov{f}}$.
\par
We follow Ref. \price\ and perturb this geometry introducing an ingoing flow of
radiation from the exterior region, which represents the radiative tail
of the gravitational collapse.
It is described by a massless
scalar field $f(v)$ with energy-momentum tensor
\eqn\infi{
T_{vv} = {{L(v)}\ov{4\pi r^2}},\ \ \ L(v)\sim v^{-p},
}
where $L(v)$ is a luminosity function and $p\ge 12$.
We note that the null coordinate $v$ is the natural advanced time coordinate
used by inertial observers at infinity 
and its value is $\infty$ both on future null infinity and on the Cauchy
horizon.
\par
An exact solution of the Einstein field equations taking into account
the source given in eq. \infi\ is known and is called the Vaidya
spacetime. The line element is written in the light-cone gauge
\eqn\vadli{
ds^2=-f(v,r)dv^2 + 2dvdr + r^2 d\Omega^2,
} 
where  
\eqn\mvai{
f=1-{{2m(v)}\ov{r}}+{{Q^2}\ov{r^2}}
}
and the mass function $m(v)$ is given by the equation
\eqn\priv{
{{dm}\ov{dv}}=4\pi r^2 T_{vv}=L(v) \sim  v^{-p}.
}
\par
The Vaidya spacetime so defined is regular at the Cauchy
horizon.
However, let us consider a geodesic observer crossing the surface
$v=\infty$ at the Cauchy horizon. From the geodesic equations 
it is easy to see that close to the horizon its velocity is such that
\eqn\obve{
\dot r \simeq -c, \ \ \ \dot v \simeq c e^{k_- v},
}
where $c$ is a positive constant
and $k_-$ is given in eq. \hori. Therefore he measures an 
energy influx given by
\eqn\inkj{
\rho_{obs}\sim T_{vv}\dot v^2 \sim v^{-p}e^{2k_- v}.
}
This quantity clearly diverges as $v=\infty$ due to the presence
of the exponential blueshift factor $e^{2k_- v}$.\foot{This is the
so called proper-time compression effect: the free falling observer 
sees in a finite amount of proper time an infinite number of ingoing
waves accumulating at the Cauchy horizon.}
\par
It is reasonable to expect that in even more realistic scenarios of
gravitational collapse a singularity will appear
 at the Cauchy horizon. The key observation
 has been made by Poisson and Israel \mainf:
a  star will in general emit also outgoing  
 gravitational waves during its collapsing phase.
 It is striking that no matter how $T_{uu}$ is, as long as
 it is nonvanishing, the combination of both fluxes will cause
 the mass function to diverge at the Cauchy horizon as
 \eqn\masd{
 m(v)\sim v^{-p}e^{k_-v}.
 }
 We note that the divergence in eq. \masd\ is 
 in some sense `milder' than 
 in eq. \inkj.
 This behaviour is called the mass inflation phenomenon. It is
 responsible for the appearance of a scalar curvature singularity
 at the Cauchy horizon 
 ($R$ has the same divergence as in eq. \masd)
 and therefore predictability of the
 Einstein field equations is saved in this model.
 \par
A different picture arises when one considers perturbations of the extremal 
Reissner-Nordstr\"om black holes \poisson. These are defined by the
condition $m_0=|Q|$ and, from eqs. \upa\ and \hori, $f=(1-{{m_0}\ov r})^2$,
$r_+=r_-=m_0$ and $k_{\pm}=0$. 
\par \n
An appropriate perturbation for this spacetime is of the type in eq. \infi\
(with $p$ numerically smaller than for the non-extreme case, but still
$p\ge 8$). The geodesic equations can be easily solved in the vicinity
of the degenerate horizon and give
\eqn\bahh{
\dot v \sim v^2, \ \ \ \dot r \sim - c.
}
The energy density measured by the free falling observer is then
\eqn\bahia{
\rho_{obs} \sim v^{-p+4}, 
}
vanishing in the limit $v\to \infty$, i.e. at the Cauchy horizon.
\par \n
In this case the Cauchy horizon is stable and the backreaction of the
influx eq. \bahia\ 
together with a generic outflux
does not affect the regularity of the geometry there.
\par
To summarize, in the space ($m_0$, $Q$) of the parameters characterizing the 
Reissner-Nordstr\"om geometries 
the perturbation analysis has shown that the Cauchy horizons are
generically unstable, except for the region of zero measure 
$m_0=|Q|$.
\par \n
It is worthwhile to stress that the unperturbed Reissner-Nordstr\"om
are, by virtue of Birkhoff's theorem, the only static, spherically
symmetric and asymptotically flat solutions of Einstein-Maxwell equations. 

\subsec{The Reissner-Nordstr\"om-De Sitter spacetime}

The line element of the Reissner-Nordstr\"om-De Sitter
geometry can be written as in eq. \ago, where now
\eqn\frds{
f=1-{{2m_0}\ov{r}}+{{Q^2}\ov{r^2}}-{{\Lambda}\ov 3} r^2
}
and $\Lambda$ is the cosmological constant.
\par
The equation $f=0$, which defines the horizons, is a quartic. We
take three of its real roots to be physical and we call them
$r_i$, $r_h$ and $r_c$, where $r_i<r_h<r_c$. 
$r=r_i$ denotes the location of the inner,
and Cauchy, horizon; $r_h$ is the outer horizon and $r_c$ is the
cosmological horizon.
The surface gravity at the horizons is defined, as usual,
by $k={1\ov 2} |\dd_r f|$ evaluated at $r=r_i$, $r=r_h$ and
$r=r_c$.
Null directions
are again $u$ and $v$, where $dv=dt+{{d\sigma}\ov{f}}$ and 
$du=dt-{{d\sigma}\ov{f}}$.
\par
The causal structure of this spacetime is represented in Fig. 2.
As opposed to the diagram of Fig. 1, now the 
causal past of the Cauchy horizon does not contain all of the external 
region, its
 boundary being represented by the cosmological horizon.
\par
To construct a perturbation appropriate for this
geometry we have to take into account the fact that the
coordinate $v$ is not inertial at $r=r_c$.
A finite perturbation in the local inertial frame there
 once transformed to $v$ coordinate reads \bradpoi\
\eqn\fipi{
T_{vv}={{Ke^{-2k_c v}}\ov{4\pi r^2}},
}
where $K$ is a constant and $k_c$ is the surface gravity at the
cosmological horizon.
\par
An exact solution of the Einstein field equations taking into
account the contribution of eq. \fipi\ is the 
Vaidya-Reissner-Nordstr\"om-De Sitter spacetime. Its line element is of the
form in eq. \vadli, $f(r,v)$ is given by an equation like
\frds\ with $m=m(v)$.
\par
Despite the regularity of the geometry at the Cauchy horizon, a free
falling observer in its vicinity will measure an influx of
energy given by
\eqn\vigh{
\rho_{obs}\sim e^{2(k_i-k_c)v},
}
where $k_i$ is the surface gravity associated with the Cauchy horizon.
It diverges provided $k_i>k_c$.
\par
We can proceed as in the case of the Reissner-Nordstr\"om spacetime,
i.e. we couple the influx of matter with generic outflowing null radiation
crossing the Cauchy horizon.
The behaviour of the mass function in the limit $v=\infty$
has been computed in Ref. \branusi\ and the results show that
\eqn\mard{
m(v) \sim e^{(k_i - 2k_c)v}.
}
\par
From these results we can identify three different regions
in parameter space ($m_0$, $Q$, $\Lambda$):
\par \n
i) $k_i\leq k_c$, the Cauchy horizon is completely stable;
\par \n
ii) $k_c<k_i\leq 2k_c$, divergent influx of radiation at $r_i$,
but no mass inflation; \foot{Although the
mass function at the Cauchy horizon is finite, in \branusi\ it is shown
that the 4d metric has the Kretschmann invariant
$R^{\alpha\beta\gamma\delta}R_{\alpha\beta\gamma\delta}$
divergent there.}
\par \n
iii) $k_i >2k_c$, both divergent influx and mass inflation.
\par
Therefore, this simple model of crossflowing streams of null matter
shows that in general when we have non asymptotically
flat spacetimes the traversibility of the Cauchy horizon 
can be physically possible. To analitically
extend the metric beyond it we need to impose boundary
conditions on the timelike singularities represented
in Fig. 2. In particular, the diagram of Fig. 2 was obtained
from the assumption that `nothing escapes from the singularity'.
\par
We note finally that in obtaining the results in eqs.   
\vigh\ and \mard\ the only features we used of the spacetime 
in Fig. 2 are the surface gravities at the Cauchy and cosmological
horizons. We did not take
into any account the structure of the geometry beyond the
cosmological horizon.

\newsec{Two-dimensional black holes in accelerated frames}

In this section we simply recall the form of the classical black hole
solutions of the models introduced in Ref. \fabru\ and consider briefly,
for the purposes of the present paper,
 their
causal structure.
\par
We consider the action $S_{cl}=S_n +S_M + S_{EM}$, $S_n$ given in
eq. \accla,
\eqn\acmat{
S_M = - {1\ov {4\pi}} \sum_{i=1}^{N}\int d^2 x \sqrt{-g}
(\nabla f_i )^2 \  
}
 and (see Ref. \fabal)
\eqn\azem{
S_{EM}={1\ov {2\pi}}\int d^2 x \sqrt{-g}e^{{{2n-4}\ov{n}}\phi}(-2F_{\mu\nu}
F^{\mu\nu}),
} 
where $F_{\mu\nu}$ is the e.m. field tensor. 
\par
The equations of motion derived from varying $S_{cl}$ with respect 
to $\phi$ and $g_{\mu\nu}$ are
\eqn\motobb{{R\ov{n}}-{4\ov{n^2}}(\nabla\phi)^2+{4\ov n}\nabla^2\phi+
4\lambda^2e^{{{2-2n}\ov{n}}\phi}+{{2n-4}\ov{n}}
e^{{{2n-2}\ov{n}}\phi}F^{\mu\nu}F_{\mu\nu}=0 \ ,}
$$
g_{\mu\nu}[{4\ov n}(-{1\ov 2}+ {1\ov n})(\nabla\phi)^2 -{2\ov n}
\nabla^2\phi +
e^{ {{2n-2}\ov{n}}\phi}
F^{\mu\nu}F_{\mu\nu}-2\lambda^2 e^{{{2-2n}\ov{n}}\phi}]+
{4\ov n}(1-{1\ov n})\partial_{\mu}\phi\partial_{\nu}\phi -$$
\eqn\motoaa{
-4e^{{{2n-2}\ov{n}}\phi}
F_{\mu\alpha}F_{\nu}^{\alpha}+{2\ov n}\nabla_{\mu}\partial_{\nu}\phi+
e^{{2\ov n}\phi}T_{\mu\nu}^M =0\ .
}
\par
The equations of motion of $F_{\mu\nu}$ are easily derived from $S_{EM}$:
\eqn\eqem{\nabla_{\nu} (e^{{{2n-4}\ov{n}}\phi}F^{\mu\nu})=0\ . }
The e.m. field tensor is totally antisymmetric and so can be written as
$F_{\mu\nu}=Fe_{\mu\nu}$, where $e_{\mu\nu}=e_{[\mu\nu]}$ and $e_{01}=
\sqrt{-g}$. Inserting this into \eqem\ we get the solution
$F_{\mu\nu}=Qe^{{{4-2n}\ov{n}}\phi}e_{\mu\nu}$, where $Q$ is a constant representing
the charge on the black hole.
\par 
The static black hole solutions can be easily derived in the
`Schwarzschild-Rindler' gauge (for the details see Ref. \fabal)
 ($\sigma$, $t$), where
\eqn\schw{ds^2=e^{2(1-n)\lambda\sigma}[-fdt^2 + {1\ov f}d\sigma^2 ],
\ \ \ \phi=-n\lambda\sigma ,
}
and $f$ is defined by 
\eqn\frn{
f=1-{{2m_0}\ov{\lambda}}e^{{2\ov n}\phi}+
{{Q^2}\ov{\lambda^2}}e^{{4\ov n}\phi}.
}
\par
Consider first the case $n<1$. For
$m_0 > |Q|$ 
 we have two Killing horizons located at 
$\sigma=\sigma_{\pm}$, where
\eqn\hori{
e^{2\lambda\sigma_{\pm}}=e^{-{2\ov n}\phi_{\pm}}=
{{(m_0\pm \sqrt{m_0^2-Q^2})}\ov{\lambda}}\ .}
$\sigma_+$ is the location of the event horizon and
$\sigma_-$ is the inner, and Cauchy, horizon.
The causal diagram is the same as that in Fig. 1.
The asymptotic region is defined by $\sigma=\infty$ and note that there
the metric is not Minkowskian, but Rindlerian, due to the presence
of the conformal factor $e^{2(1-n)\l\sigma}$. 
\par
We turn now to the case $n>1$. Concerning the global causal structure
of these spacetimes, we still have Cauchy and black hole horizons at
$\sigma=\sigma_{\pm}$ as in eq. \hori.
However, the region $\sigma=\infty$ 
doesn't represent anymore the asymptotic region, but is the location
of another horizon, the acceleration horizon, where the line element takes
the Rindler form. The metric can be analitically extended across this
horizon and the results of this analysis are presented in Ref. \balfab,
where it is shown that a true asymptotic region doesn't
exist in these spacetimes.
\par
However, as noted at the end of section 2.2, all we need to discuss
the classical stability and the mass inflation phenomenon at the Cauchy horizon
is the causal structure of the spacetime in its causal past.
Therefore for our purposes the geometries with $n<1$  
have the same features of standard Reissner-Nordstr\"om spacetime
and those for $n>1$ are similar to the Reissner-Nordstr\"om-De Sitter
geometry.
\par
A useful quantity is the local
surface gravity $k$, defined by (see for example \wald)
\eqn\surgra{
\k={1\ov 2}\big| {{\dd_{\sigma} g_{tt}}
\ov{\sqrt{-g_{\sigma\sigma}g_{tt}}}} \big| =
\big| \l (1-n) f + {1\ov 2} f_{,\sigma} \big|\ .
}
One finds that at $\sigma_{\pm}$ 
it is 
\eqn\sagra{
\k_{\pm} = {1\ov 2}|f_{,\sigma_{\pm}}|
= \l e^{-2\l\sigma_{\pm}} (e^{2\l\sigma_+}
 - e^{2\l\sigma_- }).
}
 Note that, provided $m_0>|Q|$, it is always $k_- >k_+$.
In the case $n>1$ at the acceleration horizon 
\eqn\grac{
\k_{ah}\equiv \k (\sigma=\infty) =
\l (n-1).
}
\par
Since in the following we will be interested 
also in solutions arising from
purely incoming massless matter we write down the metric 
and the dilaton in the chiral
gauge 
\eqn\lico{ds^2=e^{2(1-n)\lambda\sigma}(-fdv^2 +2dvd\sigma ),
\ \ \ \phi=-n\l\sigma,
}
where
\eqn\lkla{
f=1-{{2m(v)}\ov{\lambda}}e^{{2\ov n}\phi}+
{{Q^2}\ov{\lambda^2}}e^{{4\ov n}\phi}.
}
Allowing for purely ingoing $f_i$-wave $f_i=f_i(v)$ 
the $vv$ component of eq. \motoaa\
gives the additional condition 
\eqn\massa{
{{dm}\ov{dv}}={1\ov 4}\sum (\partial_v f_i)^2.
}
For $f_i\neq 0$ the solution resembles,
apart from the conformal factor $e^{2(1-n)\l\sigma}$,
the Vaidya solution of General Relativity.
 In general $m$ is a function of $v$ and we will refer to the scalar $m$
in 
\eqn\nabi{
(\nabla\phi)^2=n^2\lambda^2e^{-2(1-n)\lambda\sigma}(1-{{2m}\ov{\lambda}}
e^{-2\lambda\sigma}+{{Q^2}\ov{\lambda^2}}e^{-4\lambda\sigma})
}
as the mass function of the black hole.

\newsec{Cauchy horizon stability and mass inflation in the case $n<1$
(asymptotically flat space-times)}

The interesting and crucial thing to note about the solution 
in eq. \schw\ is that
it is static as referred to the asymptotic Rindler observers.
An inertial observer at infinity will, in fact,
detect a 
time dependent gravitational field. The inertial frame there is defined
by the coordinates 
\eqn\mink{
\l y^+ = {{e^{\l (1-n) v}}\ov{(1-n)}},
}
\eqn\minku{
-\l y^- = {{e^{-\l (1-n)u}}\ov{(1-n)}}. 
}
Neglecting the term proportional to $Q$, which is subleading at infinity
as compared to that proportional to $m_0$, 
in this frame the metric takes the form
\eqn\memi{
ds^2  = -f^n (y^+, y^-)dy^+ dy^-
=-{{dy^+dy^-}\ov{\big[ 1+{{2m_0}\ov{\l}}\big( -\l^2 (1-n)^2 y^+y^- 
\big) ^{{1\ov{n-1}}} \big] ^n }}
\  .
}
Note that in the external region
 $y^+ \in\ [0,+\infty [$ and $y^- \in\ ]-\infty , 0]$.

In the following we will use the `natural' 
frame of the accelerated observers, comoving with the black hole.

\subsec{Stability of the Cauchy horizon}

In order to analyse the stability of the Cauchy horizon, we must consider 
gravitational perturbations generated in the external (asymptotically
flat) region. Following the discussion of section I, we model them
 in the form of ingoing
null matter flowing into the black hole.
Such perturbations 
decay with
an inverse power law in the advanced time of the asymptotic inertial
observer and are described by the energy-momentum tensor 
\eqn\cipi{
T_{++}={1\ov 2} ({{df}\ov{dy^+}})^2=
2\gamma (\lambda y^+)^{-p}, 
}
$\gamma$ being a constant and 
 $p\ge 12$. 
Transforming to $v$ coordinate we get
\eqn\tipu{
T_{vv}=2\gamma (1-n)^p e^{\lambda (1-n) (-p+2)v}.
} 
Provided that the influx is turned on at a finite $v_0$  
the mass function from eq. \massa\
is given by
\eqn\mass{m(v)=m_0 - 
{{\gamma (1-n)^{p-1}}\ov{\lambda (p-2)}}e^{\lambda (1-n)(-p+2)v}
\ . }
This simply tells us that in the external region the 
black hole settles down to the final mass $m_0$. 
\par
Let us now consider a free falling observer close to the Cauchy horizon,
defined again by $v=\infty$. Solving the geodesic equations  we get,
as in section 2,
$\dot v \sim e^{k_- v}$
 and $\dot \sigma =\cst$, where a dot means derivative with respect to
the proper time. 
Thus the energy density he measures is 
\eqn\deen{\rho_{obs}\sim T_{vv}\dot v ^2\sim e^{\lambda (1-n)(-p+2)v+2k_- v}\ . }
This quantity remains finite whenever 
\eqn\stab{
(1-n)(-p+2)\l +2k_- \leq 0\ .
}
We can use eqs. \hori\ and \sagra\ to rewrite eq. \stab\
in the form
\eqn\mocl{
 m_0^2\le Q^2 {{[4+(1-n)(p-2)]^2}\ov{16+8(1-n)(p-2)}}
}
(together with $m_0^2 \ge Q^2$).
\par Physically, this result is due to the fact that the blueshift factor,
which is exponential in the null time coordinate $v$ in the
Reissner-Nordstr\"om spacetime, now has a power 
law behaviour in $y^+$ (this is easily seen writing eq. \deen\
 in terms of $y^+$).\foot{Therefore, what makes the Cauchy horizon
 unstable is not just the infinite accumulation of ingoing waves
 in its vicinity, but also the `strengh' of this accumulation, i.e.
 the behaviour of the blueshift factor.}
The same thing happens for the extreme Reissner-Nordstr\"om black holes,
see eqs. \bahh\ and \bahia, where $k_-=0$.\foot{The stability 
of other 2d dilaton black holes with $k_-=0$ at the Cauchy horizon
is considered in Ref. \chma.}

\subsec{Mass inflation}

Our aim is to construct an approximate solution to the equations of motion
in the presence both of incoming and outgoing fluxes of null radiation
close to the Cauchy horizon.
It is useful to write down eqs. \motobb\ and \motoaa\
 in the conformal
gauge defined by
\eqn\kjss{
ds^2=-e^{2\rho}dx^+dx^- \equiv -Fe^{2\phi}dx^+dx^-,
}  
where we have introduced the quantity $F\equiv e^{2(\rho -\phi)}$. 
It is
\eqn\como{\dd_+\dd_-(\ln F)= -2Q^2 Fe^{{6\ov n}\phi}\ , }
\eqn\comb{\dd_+\dd_-(e^{-{2\ov n}\phi})=(-\lambda^2 + Q^2 
e^{{4\ov n}\phi} ) F \ ,}
while the constraints are
\eqn\costr{\dd_{\pm}^2(e^{-{2\ov n}\phi})- 
\dd_{\pm}(e^{-{2\ov n}\phi})\dd_{\pm}(\ln F)=
-{1\ov 2}\sum_{i=1}^{N}(\partial_{\pm} f)^2 \ . }
\par
For simplicity let us imagine
that the inflow is turned on at a finite advanced
time $x^+=x_0^+$ and that
the outflow, crossing the Cauchy horizon, starts at $x^-=x^-_0$ (see Fig. 3).
\par
The spacetime is then divided in different zones: a pure inflow
and a pure outflow regions, whose geometries are correctly described by the
Vaidya type metric in eq. \lico\
(in the outflow case $v$ is replaced by $u$ and \lico\ remains the same
apart from a minus sign in $dud\sigma$),  
static sectors where the metric takes the form eq. \schw\ and finally
the most interesting one where both fluxes are present.
\par
We concentrate our discussion to this last region, close to $x_0^-$, and
choose $x^+$ such that $x^+=0$ at the Cauchy horizon.
\par 
Note that for $x^-<x^-_0 $ and close to the Cauchy horizon the solution
is almost static
(see eq. \mass\ in the limit $v\to \infty$),
 the dilaton reaching the finite value $\phi_-$. 
We fix the $x^+$ coordinate in such a way that 
$F(x^-_0,x^+)=F(x^-,x_0^+)=e^{-2\phi_-}$ .
This ensures that $x^+$ is regularly related to the
Kruskal advanced time coordinate associated with the inner horizon, 
i.e. $\sim - e^{-k_- v}$ (see for example Ref. \balb).
\par
The form of $T_{++}^M$ then follows from eq. \tipu\ after
a coordinate change, that is 
\eqn\emmo{
T_{++}^M\sim (x^+)^{-{{\lambda (1-n)(-p+2)}
\ov{k_-}}-2},
} 
while the exact relation between $u$ and $x^-$ along
with the shape of $T_{--}^M$ ($\neq 0$) will not concern us here, since the
conclusions that we'll draw will be independent 
on them.\foot{Note that $T_{++}^M$ in eq. \emmo\ has exactly the same
divergence as $\rho_{obs}$ in eq. \deen. }
\par 
An approximate solution to 
$F$ close to $x_0^-$ and to the Cauchy horizon
following from eq. \como\ and with the above
initial conditions is
\eqn\appro{
F\simeq e^{-2\phi_-}exp \big(\  {-2Q^2 e^{({6\ov n}-2)\phi_-} 
(x^- -x^-_0) (x^+ -x^+_0)}\ \big) \ .}
This permits to solve, in the same limit, 
the constraints eqs. \costr\ along with 
eq. \comb\ giving 
$$ e^{-{2\ov n}\phi}\simeq e^{-{2\ov n}\phi}(x^+_0,x^-_0) - 
\int_{x^+_0}^{x^+}
d\xi F(\xi)\int _{x^+_0}^{\xi} 
{{d\xi^{'}}\ov{F^(\xi^{'})}}T_{++}^M(\xi^{'})-$$
\eqn\solu{
-\int_{x^-_0}^{x^-}d\xi F(\xi)\int_{x^-_0}^{\xi}
{{d\xi^{'}}\ov{F(\xi^{'})}}
T_{--}^M(\xi^{'}) 
+x^+ (-\lambda^2 e^{-2\phi_-}+Q^2e^{({4\ov n}-2)\phi_-})
(x^- -x^-_0)\ . }
\par
Note that both $F$ and the metric stay bounded as $x^+\to 0$.  
However when one calculates the
Ricci scalar $R=4F^{-1}e^{-2\phi}\partial_+\partial_- (2\phi +\ln F)$
we have a sum of different terms of which only one is potentially
divergent, namely that proportional to 
\eqn\popi{
\int_{x_0^+}^{x^+} {{d\xi}\ov{F(\xi)}}T_{++}^M(\xi) \int_{x_0^{-}}^{x^-}
{{d\xi^{'}}\ov{F^(\xi^{'})}}T^M_{--}(\xi^{'}).
}
\par
Under the assumption that $\int_{x_0^{-}}^{x^-}
{{d\xi^{'}}\ov{F^(\xi^{'})}}T^M_{--}(\xi^{'})$ is finite, we can easily
transform the other integral to $v$ coordinate and see that as $v\to \infty$
it behaves as 
\eqn\kkkk{
e^{\l(1-n)(-p+2)+k_- v}.
}
We stress that the dependence on $k_-$ of \kkkk\ is the same as
in eq. \masd. This is typical of mass inflation.
\par 
We get therefore the result that if 
\eqn\stabu{
\lambda (1-n)(-p+2) +k_- \leq 0
}
then $R$ is finite at the Cauchy horizon, but when this inequality
is not satisfied the scalar curvature diverges.
Eq. \stabu\ can also be rewritten as 
\eqn\mmmm{
m_0^2\le Q^2 {{[2+(1-n)(p-2)]^2}\ov{4+4(1-n)(p-2)}}.
}
\par
Therefore, also at this level the possibility of having a regular Cauchy
horizon is not completely ruled out. 
\par
According to the results in eqs \stab\ and \stabu\ in these two-dimensional
theories three different regimes can be identified:
\par \n
i) $k_- \leq {{\l(1-n)(p-2)}\ov 2}$, complete stability;
\par \n
ii) ${{\l(1-n)(p-2)}\ov 2}< k_- \leq \l(1-n)(p-2)$, $\rho_{obs}$ diverges
but the spacetime remains regular;
\par \n
iii) $k_- > \l(1-n)(p-2)$, both $\rho_{obs}$ and $R$ are unbounded 
at the Cauchy horizon.
\par
A comparison with the analysis of perturbations carried out in the
Reissner-Nordstr\"om  geometry (section 2.1) shows that now 
complete stability of the Cauchy horizon is achieved 
in a region of nonzero measure in the space of the parameters ($m_0$, $Q$,
$\l$, $n$) (in the plane ($m_0^2$, $Q^2$) it is the dotted region of Fig. 4).   
\par \n 
This is a new result concerning asymptotically flat spacetimes.

\newsec{Stability analysis for $n>1$}

We have already noted that this case is qualitatively similar to that
considered in section 2.2, concerning the Reissner-Nordstr\"om-De Sitter
spacetime.
As considered there, we introduce a null perturbation in the spacetime
which is regular at the acceleration horizon and write 
\eqn\prtr{
T_{vv}=Ke^{-2k_{ah}v},
}
where $K$ is a constant and $k_{ah}$ is given by eq. \grac.
Note that now the coordinates in eqs. \mink\ and \minku\ 
define the Kruskal frame, regular at the acceleration horizon.
\par
Simple stability arguments at the Cauchy horizon, as in the
previous section, show that geodesic observers there will measure an energy
density given by
\eqn\chhh{
\rho_{obs}\sim e^{2(k_- - k_{ah})v}.
}
As $v\to\infty$ the stability
is guaranteed when the inequality
\eqn\ghas{
k_-\leq k_{ah}
}
is satisfied.
By making use of the eqs. \sagra\ and \grac\ we can rewrite \ghas\ as
\eqn\moel{
m_0^2 \leq Q^2 {{(1+n)^2}\ov{4n}}.
}
The insertion of a nonvanishing outflux of null radiation modifies the
picture more or less as in section 4.2.
Here we don't write down the explicit calculations, 
but just limit ourselves
to note that as a consequence the curvature scalar $R$ at the Cauchy horizon
goes as
\eqn\cosc{
R \sim e^{(k_- - 2k_{ah})v}.
}
This implies that the resulting spacetime is regular at the Cauchy horizon
only when
\eqn\fist{
k_- \leq 2k_{ah},
}
which can be rewritten more elegantly as
\eqn\cgbq{
m_0^2 \leq Q^2 {{n^2}\ov{2n-1}}.
}
The formulas in eqs. \ghas\ and \fist\ are very similar to those derived in
section 2.2, provided we replace $k_{ah}$ with $k_c$.
Therefore all the conclusions presented there on the stability
of the Cauchy horizon are valid here too. 

\newsec{Quantum effects, backreaction and conclusions}
In this paper we have shown that the instability of the Cauchy horizon, 
peculiar to the 
asymptotically flat solutions of General Relativity,
is not generic in two-dimensional theories of gravity.
\par
Indeed, we presented a class of dilaton-gravity models 
for which, due to the asymptotic Rindlerian form of the metric,
the Cauchy horizon can be stable under generic small perturbations
generated in the external universe and, also, the scalar curvature
stays regular there.
\par
This suggests the idea that, in principle, an observer falling into
the black hole is not destroyed at the spacetime singularity
but, due to the timelike `repulsive' character of the singularity itself, 
can safely travel through the black hole tunnel of Fig. 1 and then emerge
into another asymptotically flat region. 
\par
It is interesting to investigate, at this point, whether quantum
effects can modify the classical picture emerged in this paper.
In Ref. \fabal\ we have computed the v.e.v. of the energy-momentum tensor
of $N$ massless conformally coupled scalar fields in these
spacetimes. Now we wish to determine their backreaction on the geometry
close to the Cauchy horizon.
\par
The semiclassical theory that was considered in \fabru,
which keeps into account the effects of the
Hawking radiation,
is described by the action   
\eqn\acqua{
S=S_n +S_{M}+S_{EM} 
+{{\k}\ov{2\pi}}\int d^2 x \sqrt{-g}[{{1-2n}\ov{2n}}\phi R + {{n-1}\ov n}
(\nabla\phi)^2 - {1\ov 4} R (\nabla ^2)^{-1} R] \ ,
}
where $\k={N\ov {12}}$ and 
$S_n$, $S_M$ and $S_{EM}$ are considered in \accla, \acmat\ and \azem. 
\par 
Let's perform the field redefinitions
\eqn\fired{ \chi = \k \rho + ({1\ov{2n}}-1)\k \phi + e^{-{2\ov n}\phi}, }
\eqn\firb{\Omega= {{\k}\ov{2n}}\phi + e^{-{2\ov n}\phi}\  }
and work in the conformal gauge.
If $Q=0$ the action \acqua\ can be cast in the `free field' form 
\eqn\liou{
S={1\ov{\pi}} \int d^2 x [{1\ov {\k}}(-\dd_+\chi\dd_-\chi +\dd_+\Omega
\dd_-\Omega) + \l^2 e^{{2\ov{\k}}(\chi - \Omega)}] + {1\ov 2}
\sum_{i=1}^{N}\dd_+f_i\dd_-f_i ]
}
and the models become exactly solvable.
For $Q\neq 0$ it is no more possible to write down explicitly the action,
not knowing the inverse transformation $\phi=\phi(\chi, \Omega)$.
However we can still derive the equations of motion, i.e.
\eqn\eqcar{ \dd_+\dd_-\chi = (-\l^2 + Q^2 e^{{4\ov n}\phi})
e^{{2\ov{{\k}}}(\chi - \Omega)}\ , }
\eqn\eqcra{ \dd_+\dd_-(\chi - \Omega )=  {2k \ov n}  Q^2 {{e^{{4\ov 
n}\phi}}\ov {\Omega^{'}}} e^{{2\ov{\k}}(\chi - \Omega)\  } }
and the constraint equations
\eqn\cotra{
\k t_{\pm}={1\ov {\k}}(-\dd_{\pm}\chi\dd_{\pm}\chi + \dd_{\pm}\Omega
\dd_{\pm}\Omega ) + \dd_{\pm}^2 \chi + {1\ov 2} \sum _{i=1}^{N}
\dd_{\pm}f_i\dd_{\pm}f_i. }
\par
Here the choice of $t_+$ becomes crucial. 
Physically, it measures how the quantum 
state of the $f_i$ fields is related to the vacuum defined
in terms of the modes $x^{\pm}$, regular at the Cauchy horizon. 
\par
We could require, as in \balb,
physical conditions such that the black hole is in thermal equilibrium
with a heat bath in order to isolate effects on the interior from those of
the evaporation. Or, as we did in \fabru, we could demand that
the state of the $f_i$ fields is the one naturally associated
to Rindler modes $e^{-iwu}, e^{-iwv}$.
\par
In both cases we find that  
\eqn\tpiu{
t_+(x^+)\sim {{\alpha}\ov{{x^{+2}}}}, 
}
where $\alpha$ is a 
positive constant (this is easily seen by performing the Schwarzian derivative
between $e^{k_+ v}=(-x^+)^{-{{k_+}\ov{k_-}}}$ (in the first case) or 
$v=-{1\ov{k_-}}\ln (-k_-x^+ )$ (in the second case) and  $x^+$).
The explicit form of both $t_-(x^-)$ and $T_{--}^M$ is again not 
important for our discussion. 
\par 
Much care is needed in order to determine whether or not the r.h.s.
of eq. \eqcra\ diverges before the inner horizon is encountered. Classically,
there's a range of values of $m_0$, $Q$ and $\l$ such that it is always 
$e^{-{2\ov n}\phi}>e^{-{2\ov n}\phi_{cr}}={{\k}\ov 4}$ 
(and so $\Omega^{'}\neq 0$) as $\sigma\geq\sigma_-$.
Moreover at the quantum level the $t_+$ considered in eq. \tpiu\
is equivalent to introduce a negative energy flux along the Cauchy horizon,
thus providing a `defocussing' effect that is likely to guarantee 
that $\Omega^{'}\neq 0$ during the entire evolution up to $x^+\to 0$.
In fact it is easy to show (similarly to \balb) 
that, as $x^+\to 0$, we are again 
in a weak coupling region and (only finite terms are written)
\eqn\solqu{
\chi \sim \Omega \sim -\alpha\k \ln (-k_- x^+) + \int^{x^-}d\xi
\int^{\xi}d\xi^{'}[kt_-(\xi^{'})- T_{--}^M(\xi^{'})] 
}
($T_{++}^M$ of eq. \emmo\ does not contribute in this limit).
The Ricci scalar in the same limit diverges and behaves as
\eqn\ricsca{
R \sim n e^{({4\ov n}-2)\phi}{1\ov{x^+}}\int^{x^-}d\xi [T_{--}^M(\xi)-
\k t_-(\xi)]\ .
}
\par
Therefore while at the classical level we could always choose our 
parameters $n$ and $k_-$ such that the spacetime is regular at the
Cauchy horizon, quantum-mechanically this is no more possible and 
a singularity always forms independently on the value of $n$. 
\par 
We can say that in this way quantum mechanics has helped us to restore
the full predictive power of the gravitational field equations
that had been lost at the classical level.

\bigskip \bigskip
\noindent $\underline {\rm Acknowledgements}$: We thank D. Amati 
and R. Balbinot for help and useful suggestions.

\listrefs 
   
\vfill \eject
{
 \epsfxsize=6cm \epsfysize=8cm 
 \epsfbox{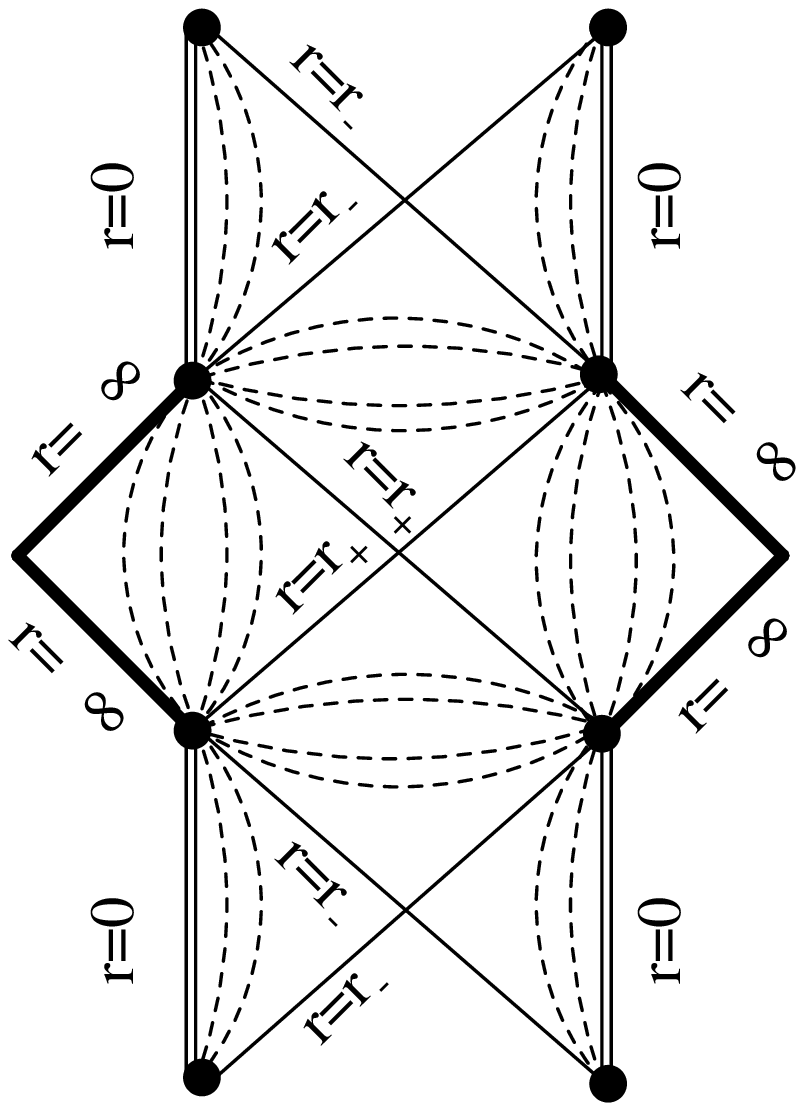}
 }
{\bf {Fig. 1:}}{
Penrose diagram of the Reissner-Nordstr\"om spacetime for $m_0>|Q|$. 
Double lines represent the singularity, 
dashed lines the curves $r=\cst$, regular lines the horizons and 
thick lines the asymptotic region.}

{
\epsfxsize=7cm \epsfysize=8cm 
\epsfbox{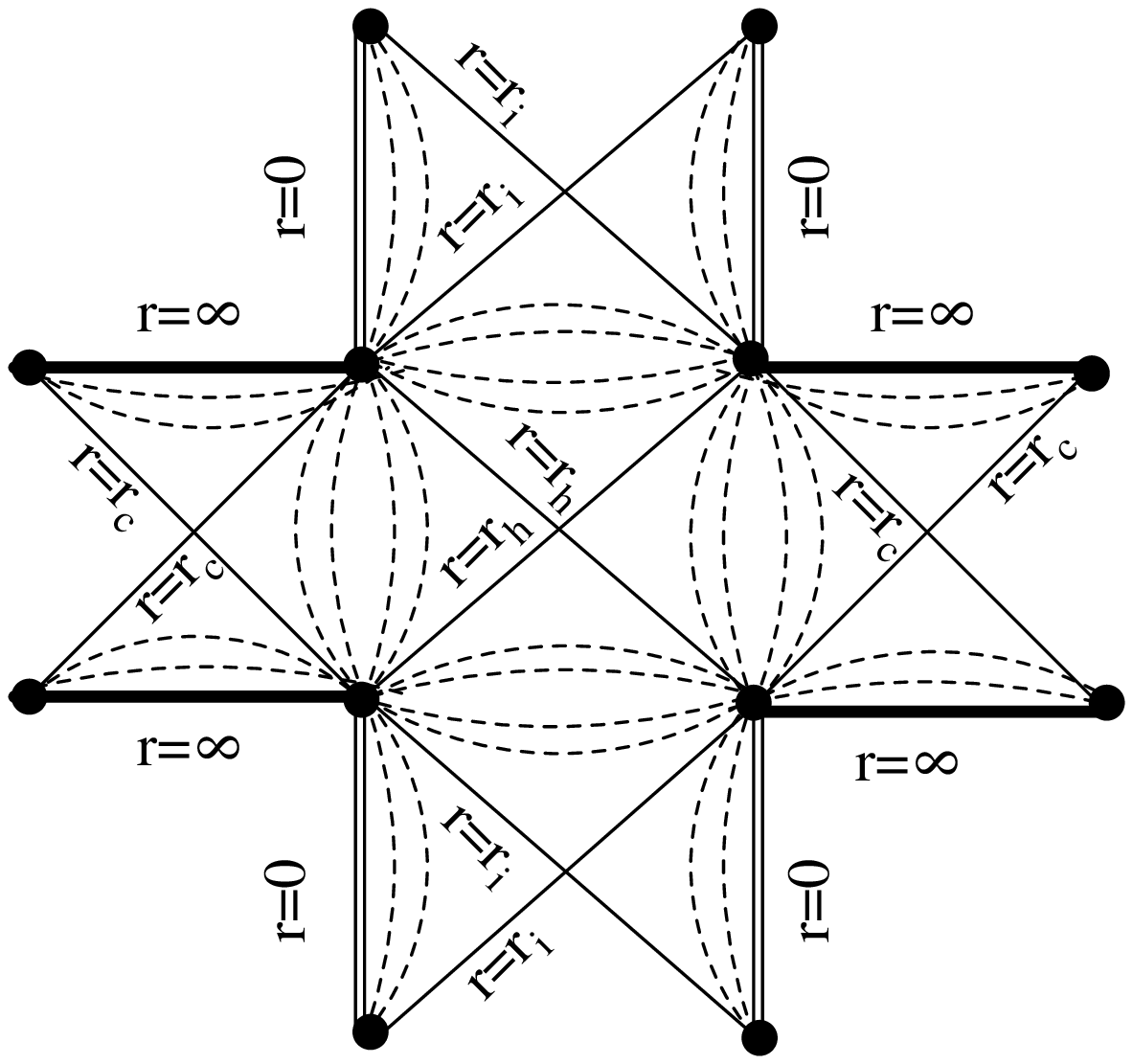}}

{\bf {Fig. 2:}}{ 
Causal structure of the Reissner-Nordstr\"om-De Sitter geometry.}

\vfill\eject
{
\epsfxsize=6cm \epsfysize=6cm 
\epsfbox{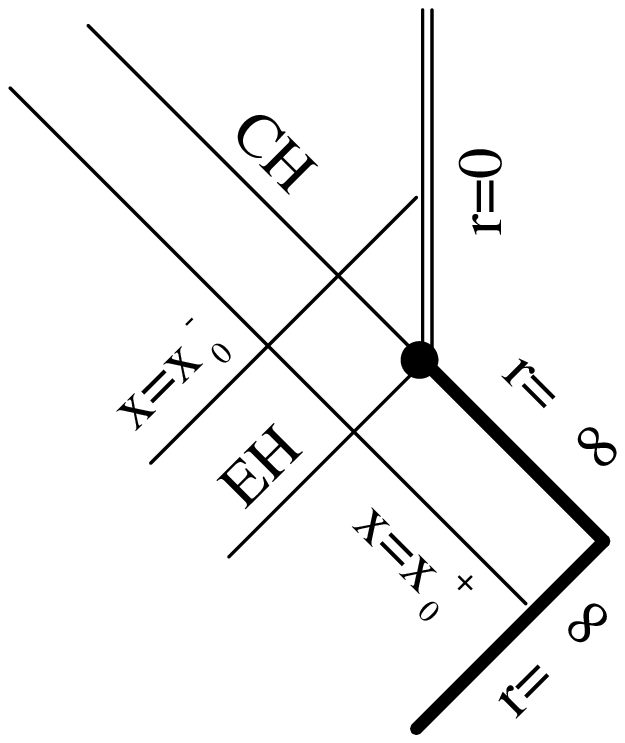}}

{\bf {Fig. 3:}}{ 
Perturbed model: the inflow starts at $x_0^+$ and the outflow at $x_0^-$.
$CH$ stands for the Cauchy horizon and $EH$ is the event horizon.}

{
\epsfxsize=6cm \epsfysize=6cm 
\epsfbox{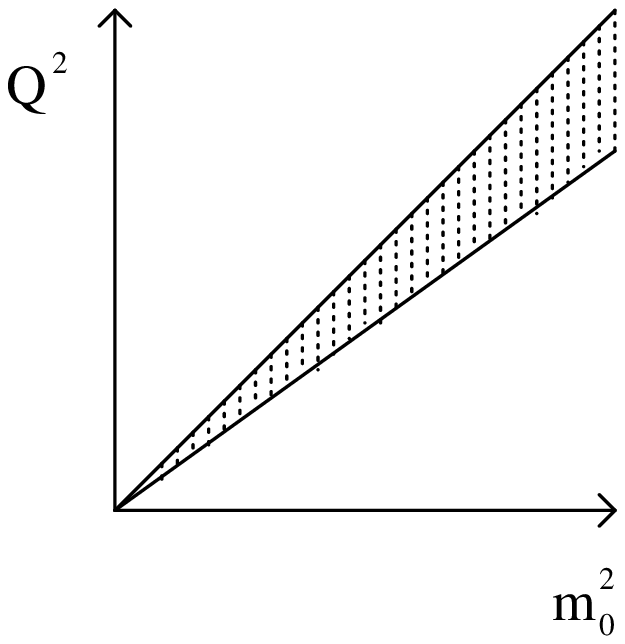}}

{\bf {Fig. 4:}}{ 
The dotted region represents the part of the ($m_0^2$, $Q^2$) plane where
the Cauchy horizon is completely stable (here we chose $n={1\ov 2}$, 
$p=12$).}

\vfill\eject

\end